\documentstyle[12pt]{article}
\newcommand{\lpar}{\stackrel{\leftarrow}{\partial}}
\newcommand{\rpar}{\stackrel{\rightarrow}{\partial}}
\begin{document}
\small
\renewcommand{\thefootnote}{\fnsymbol{footnote}}
\begin{center}
{\large\bf Linear Odd Poisson Bracket  \\
on Grassmann Algebra\\ }
\vspace{0.5cm} Vyacheslav A. Soroka\footnote{E-mail: vsoroka@kipt.kharkov.ua}
\vspace{0.5cm}\\
{\it Kharkov Institute of Physics and Technology}\\
{\it 310108 Kharkov, Ukraine}\\
\vspace{0.5cm}
\end{center}
\begin{abstract}
A linear odd Poisson bracket realized
solely in terms of Grassmann variables is suggested. It is revealed
that with the bracket, corresponding to a semi-simple Lie group,
both a Grassmann-odd Casimir function and invariant (with respect to
this group) nilpotent differential operators of the first, second
and third orders are naturally related and enter
into a finite-dimensional Lie superalgebra.
\end{abstract}
\renewcommand{\thefootnote}{\arabic{footnote}}
\setcounter{footnote}0
\bigskip

{\bf 1.\/} Recently a linear degenerate odd Poisson bracket built only of
Grassmann variables has been introduced \cite{S}. It was constructed for
this bracket, in contrast with the non-degenerate odd bracket having the
only Grassmann-odd nilpotent differential $\Delta$-operator of the second
order, at once three Grassmann-odd nilpotent $\Delta$-like
differential operators of the first, the second and the third orders with
respect to Grassmann derivatives. It was also shown that these
$\Delta$-like operators together with a Grassmann-odd nilpotent Casimir
function of this degenerate odd bracket form a finite-dimensional Lie
superalgebra. Following to \cite{S1}, in the present report we extend the
above-mentioned results to the case of an arbitrary linear odd Poisson
bracket, which is also realized solely in terms of the Grassmann variables
and corresponds to a semi-simple Lie group.

{\bf 2.\/} There is a well-known linear even Poisson bracket given
in terms of the commuting (Grassmann-even) variables $X_\alpha$
$(g(X_\alpha) = 0)$
$$
\{X_\alpha, X_\beta \}_0 = {c_{\alpha\beta}}^\gamma X_\gamma\ ,\qquad
(\alpha,\beta,\gamma = 1,...,N),
\eqno {(1)}
$$
where ${c_{\alpha\beta}}^\gamma$ are Grassmann-even
$(g({c_{\alpha\beta}}^\gamma) = 0)$ constants which, because of the main
properties of the even Poisson bracket:
$$
\{ A, B + C \}_0 = \{ A, B \}_0 + \{ A, C \}_0\ ,
\eqno {(2)}
$$
$$
g(\{ A, B \}_0) = g(A) + g(B) \pmod 2\ ,
\eqno {(3)}
$$
$$
\{ A , B \}_0 = -(-1) ^{g(A) g(B)} \{ B , A \}_0\ ,
\eqno {(4)}
$$
$$
\sum_{(ABC)}(-1) ^{g(A) g(C)} \{ A , \{ B , C \}_0 \}_0 = 0\ ,
\eqno {(5)}
$$
$$
\{ A , B C \}_0 = \{ A , B \}_0\ C +
(- 1) ^{g(A) g(B)}\ B \{ A , C \}_0\ ,
\eqno {(6)}
$$
are antisymmetric in the two lower indices
$$
{c_{\alpha\beta}}^\gamma = - {c_{\beta\alpha}}^\gamma
\eqno {(7)}
$$
and obey the conditions
$$
{c_{\alpha\lambda}}^\delta {c_{\beta\gamma}}^\lambda +
{c_{\beta\lambda}}^\delta {c_{\gamma\alpha}}^\lambda +
{c_{\gamma\lambda}}^\delta {c_{\alpha\beta}}^\lambda = 0\ .
\eqno {(8)}
$$
A sum with the symbol $(ABC)$ in (5) means a summation over cyclic
permutations of the quantities $A, B, C$. In relations (2)-(6) $A, B, C$
are functions of the variables $X_\alpha$ and $g(A)$ is a Grassmann parity
of the quantity $A$. The linear even bracket (1) plays a very important
role in the theory of Lie groups, Lie algebras, their representations and
applications (see, for example, \cite{b,km}). The bracket (1) can be
realized in a canonical even Poisson bracket
$$
\{ A , B \}_0 = A \sum_{\alpha = 1}^N
\left(\lpar_{q^\alpha}
\rpar_{p_\alpha} -
\lpar_{p_\alpha}
\rpar_{q^\alpha}\right) B\ ,
$$
where $\lpar$ and $\rpar$ are the right and left derivatives and
$\partial_{x^A} \equiv {\partial \over {\partial x^A}}$, on the following
bilinear functions of coordinates $q^\alpha$ and momenta $p_\alpha$
$$
X_\alpha = {c_{\alpha\beta}}^\gamma q^\beta p_\gamma
$$
if ${c_{\alpha\beta}}^\gamma$ satisfy the conditions (7), (8)
for the structure constants of a Lie group.

As in the Lie algebra case, we can define a symmetric Cartan-Killing
tensor
$$
g_{\alpha\beta} = g_{\beta\alpha} =
{c_{\alpha\gamma}}^\lambda {c_{\beta\lambda}}^\gamma
\eqno {(9)}
$$
and verify with the use of relations (8) an anti-symmetry property of a
tensor
$$
c_{\alpha\beta\gamma} = {c_{\alpha\beta}}^\delta g_{\delta\gamma} =
- c_{\alpha\gamma\beta}\ .
\eqno {(10)}
$$
By assuming that the Cartan-Killing metric tensor is non-degenerate
$det(g_{\alpha\beta}) \neq 0$ (this case corresponds to the semi-simple
Lie group), we can define an inverse tensor $g^{\alpha\beta}$
$$
g^{\alpha\beta}g_{\beta\gamma} = \delta^\alpha_\gamma\ ,
\eqno {(11)}
$$
with the help of which we are able to build a quantity
$$
C = X_\alpha X_\beta g^{\alpha\beta}\ ,
$$
that, in consequence of relation (10), is for the bracket (1) a Casimir
function which annihilates the bracket (1) and is an invariant of the Lie
group with the structure constants ${c_{\alpha\beta}}^\gamma$ and the
generators $T_\alpha$
$$
\{ X_\alpha, C \}_0 =
{c_{\alpha\beta}}^\gamma X_\gamma \partial_{X_\beta} C =
T_\alpha C = 0\ .
$$

{\bf 3.\/} Now let us replace in expression (1) the commuting variables
$X_\alpha$ by Grassmann variables $\Theta_\alpha$ $(g(\Theta_\alpha) = 1)$.
Then we obtain a binary composition
$$
\{\Theta_\alpha, \Theta_\beta \}_1 =
{c_{\alpha\beta}}^\gamma \Theta_\gamma\ ,
\eqno {(12)}
$$
which, due to relations (7) and (8), meets all the properties of the odd
Poisson brackets:
$$
\{ A, B + C \}_1 = \{ A, B \}_1 + \{ A, C \}_1\ ,
\eqno {(13)}
$$
$$
g(\{ A, B \}_1) = g(A) + g(B) + 1 \pmod 2\ ,
\eqno {(14)}
$$
$$
\{ A , B \}_1 = -(-1) ^{(g(A) + 1)(g(B) + 1)} \{ B , A \}_1\ ,
\eqno {(15)}
$$
$$
\sum_{(ABC)}(-1) ^{(g(A) + 1)(g(C) + 1)} \{ A , \{ B , C \}_1 \}_1 = 0\ ,
\eqno {(16)}
$$
$$
\{ A , B C \}_1 = \{ A , B \}_1\ C +
(- 1) ^{(g(A) + 1)g(B) }\ B \{ A , C \}_1\ .
\eqno {(17)}
$$
It is surprising enough that the odd bracket can be defined solely
in terms of the Grassmann variables as well as an even Martin bracket
\cite{mar}. On the following bilinear functions of canonical variables
commuting $q^\alpha$ and Grassmann $\theta_\alpha$
$$
\Theta_\alpha = {c_{\alpha\beta}}^\gamma q^\beta \theta_\gamma
$$
a canonical odd Poisson bracket
$$
\{ A , B \}_1 = A \sum_{\alpha = 1}^N
\left(\lpar_{q^\alpha}
\rpar_{\theta_\alpha} -
\lpar_{\theta_\alpha}
\rpar_{q^\alpha}\right) B
$$
is reduced to the bracket (12) providing that
${c_{\alpha\beta}}^\gamma$ obey the conditions (7), (8).

On functions $A, B$ of Grassmann variables $\Theta_\alpha$ the bracket
(12) has the form
$$
\{ A , B \}_1 = A \lpar_{\Theta_\alpha}
{c_{\alpha\beta}}^\gamma \Theta_\gamma \rpar_{\Theta_\beta} B\ ,
$$
The bracket (12) can be either degenerate or non-degenerate in the
dependence on whether the matrix ${c_{\alpha\beta}}^\gamma \Theta_\gamma$
in the indices $\alpha, \beta$ is degenerate or not. Raising and lowering
of the indices $\alpha, \beta$, the non-degenerate
metric tensors (9), (11) relate with each other the adjoint and
co-adjoint representations which are equivalent for a semi-simple Lie
group
$$
\Theta^\alpha = g^{\alpha\beta} \Theta_\beta\ ,\qquad
\partial_{\Theta^\alpha} =
g_{\alpha\beta} \partial_{\Theta_\beta}\ .
$$
Hereafter only the non-degenerate metric tensors (11) will be considered.

{\bf 4.\/} By contracting the indices in a product of the Grassmann
variables with the upper indices
and of the successive Grassmann derivatives,
respectively, with the lower indices
in (8), we obtain the relations
$$
\Theta^\alpha \Theta^\beta
( {c_{\alpha\beta}}^\lambda {c_{\lambda\gamma}}^\delta +
2 {c_{\gamma\alpha}}^\lambda {c_{\lambda\beta}}^\delta ) = 0\ ,\qquad
\Theta^\alpha \Theta^\beta \Theta^\gamma
{c_{\alpha\beta}}^\lambda {c_{\lambda\gamma}}^\delta = 0\ ,
\eqno {(18a,b)}
$$
$$
( {c_{\alpha\beta}}^\lambda {c_{\lambda\gamma}}^\delta +
2 {c_{\gamma\alpha}}^\lambda {c_{\lambda\beta}}^\delta )
\partial_{\Theta_\alpha} \partial_{\Theta_\beta} = 0\ ,\qquad
{c_{\alpha\beta}}^\lambda {c_{\lambda\gamma}}^\delta
\partial_{\Theta_\alpha} \partial_{\Theta_\beta}
\partial_{\Theta_\gamma} = 0\ ,
\eqno {(19a,b)}
$$
which will be used later on many times. In particular, taking into account
relation (18b), we can verify that the linear odd bracket (12) has the
following Grassmann-odd nilpotent Casimir function
$$
\Delta_{+3} = {1\over\sqrt{3!}} \Theta^\alpha \Theta^\beta \Theta^\gamma
c_{\alpha\beta\gamma}\ ,\qquad(\Delta_{+3})^2 = 0\ ,
\eqno {(20)}
$$
which is an invariant of the Lie group
$$
\{ \Theta_\alpha, \Delta_{+3} \}_1 =
\Theta_\gamma {c_{\alpha\beta}}^\gamma
\partial_{\Theta_\beta} \Delta_{+3} =
S_\alpha \Delta_{+3} = 0
\eqno {(21)}
$$
with the generators $S_\alpha$ obeying
the Lie algebra permutation relations\footnote{Note that below $[A, B] =
AB -BA$ and $\{ A, B \} = AB + BA$.}
$$
[ S_\alpha, S_\beta ] = {c_{\alpha\beta}}^\gamma S_\gamma\ .
\eqno {(22)}
$$

It is a well-known fact that, in contrast with the even Poisson bracket,
the non-degenerate  odd Poisson bracket has one Grassmann-odd nilpotent
differential $\Delta$-operator of the second order, in terms of which the
main equation has been formulated in the Batalin-Vilkovisky scheme
\cite{bv,bv1,blt,bt,sc,kn} for the quantization of gauge theories in
the Lagrangian approach. In a formulation of Hamiltonian dynamics
by means of the odd Poisson bracket with the help of a Grassmann-odd
Hamiltonian $\bar H$ $(g(\bar H) = 1)$
\cite{l,vpst,s,k,kn1,vst,vs,s1,s2,n,vty,s3} this
$\Delta$-operator plays also a very important role being used to
distinguish the Hamiltonian dynamical systems, for which the Liouville
theorem is valid $\Delta \bar H = 0$, from those ones,
for which this theorem takes no place $\Delta \bar H \neq 0$
\cite{S}\footnote{Note also applications of the odd bracket to the
integrability problem \cite{s4,p,ls}}.

Now let us try to build the $\Delta$-operator for the linear
odd bracket (12). It is remarkable that, in contrast with the
canonical odd Poisson bracket having the only $\Delta$-operator of the
second order, we are able to construct at once three $\Delta$-like
Grassmann-odd nilpotent operators which are differential operators of
the first, the second and the third orders respectively
$$
\Delta_{+1} = {1\over\sqrt{2}} \Theta^\alpha \Theta^\beta
{c_{\alpha\beta}}^\gamma \partial_{\Theta^\gamma}\ ,
\qquad (\Delta_{+1})^2 = 0\ ;
\eqno {(23)}
$$
$$
\Delta_{-1} = {1\over\sqrt{2}} \Theta_\gamma
{c_{\alpha\beta}}^\gamma
\partial_{\Theta_\alpha} \partial_{\Theta_\beta}
\ ,\qquad (\Delta_{-1})^2 = 0\ ;
\eqno {(24)}
$$
$$
\Delta_{-3} = {1\over\sqrt{3!}} c_{\alpha\beta\gamma}
\partial_{\Theta_\alpha} \partial_{\Theta_\beta} \partial_{\Theta_\gamma}
\ ,\qquad (\Delta_{-3})^2 = 0\ .
\eqno {(25)}
$$
The nilpotency of the operators $\Delta_{+1}$ and $\Delta_{-1}$ is a
consequence of relations (18b) and (19b). The operator $\Delta_{+1}$ is
proportional to the second term in a BRST charge
$$
Q = \Theta^\alpha G_\alpha - {1\over 2} \Theta^\alpha \Theta^\beta
{c_{\alpha\beta}}^\gamma \partial_{\Theta^\gamma}\ ,
$$
where $\Theta^\alpha$ and $\partial_{\Theta^\alpha}$ represent the
operators for ghosts and antighosts respectively. $Q$ itself will be
proportional to the operator $\Delta_{+1}$ if we take the
representation $S_\alpha$ (21) for group generators $G_\alpha$. The
operator $\Delta_{-1}$, related with the divergence of a vector field
$\{ \Theta_\alpha, A \}_1$
$$
\partial_{\Theta_\alpha} \{ \Theta_\alpha, A \}_1 =
\partial_{\Theta_\alpha} S_\alpha A =
- \sqrt{2} \Delta_{-1} A\ ,
$$
is proportional to the true $\Delta$-operator for the bracket (12).

It is also interesting to reveal that these $\Delta$-like operators
together with the Casimir function $\Delta_{+3}$ (20) are closed into the
finite-dimensional Lie superalgebra, in which the anticommuting relations
between the quantities $\Delta_\lambda$ $(\lambda = -3, -1, +1, +3 )$
(20), (23)-(25) with the nonzero right-hand side  are
$$
\{ \Delta_{-1}, \Delta_{+1} \} = Z\ ,
\eqno {(26)}
$$
$$
\{ \Delta_{-3}, \Delta_{+3} \} = N - 3Z\ ,
\eqno {(27)}
$$
where
$$
N = - c^{\alpha\beta\gamma} c_{\alpha\beta\gamma}
$$
is a number of values for the indices $\alpha, \beta, \gamma$
$(\alpha, \beta, \gamma = 1,...,N)$ and
$$
Z = D - K
\eqno {(28)}
$$
is a central element of this superalgebra
$$
[ Z, \Delta_\lambda ] = 0\ ,\qquad (\lambda = -3, -1, +1, +3)\ .
\eqno {(29)}
$$
In (28)
$$
D = \Theta^\alpha \partial_{\Theta^\alpha}
\eqno {(30)}
$$
is a "dilatation" operator for the Grassmann variables
$\Theta_\alpha$, which distinguishes the $\Delta_\lambda$-operators
with respect to their uniformity degrees in $\Theta$
$$
[ D, \Delta_\lambda ] = \lambda
\Delta_\lambda\ ,\qquad (\lambda = -3, -1, +1, +3)
\eqno {(31)}
$$
and is in fact a representation for a ghost number operator,
and the quantity $K$ has the form
$$
K = {1\over 2} \Theta^\alpha \Theta^\beta
{c_{\alpha\beta}}^\lambda c_{\lambda\gamma\delta}
\partial_{\Theta_\gamma} \partial_{\Theta_\delta}\ .
\eqno {(32)}
$$
The operator $Z$ is also a central element of the Lie superalgebra which
contains both the operators $\Delta_\lambda$ (20), (23)-(25), $Z$ (28) and
the operator $D$ (30)
$$
[Z, D] = 0\ .
\eqno {(33)}
$$

We can add to this superalgebra the generators $S_\alpha$ (21)
with the following commutation relations:
$$
[ S_\alpha, \Delta_\lambda ] = 0\ ,\qquad
[ S_\alpha, Z ] = 0\ ,\qquad [ S_\alpha, D ] = 0\ ,
\eqno {(34)}
$$
which indicate that both the Casimir function $\Delta_{+3}$ and the
operators $\Delta_\lambda$ $(\lambda = -3, -1, +1 )$, $Z$ and $D$ are
invariants of the Lie group with the generators $S_\alpha$.
In order to prove the permutation relations for the Lie superalgebra
(20)-(34), we have to use relations (18) and (19).
Note that the central element $Z$ (28) coincides with the expression for a
quadratic Casimir operator of the Lie algebra (22) for the generators
$S_\alpha$ given in the representation (21)
$$
S_\alpha S_\beta g^{\alpha\beta} = Z\ .
\eqno {(35)}
$$

{\bf 5.\/} Thus, we see that both the even and odd linear Poisson brackets
are internally inherent in the Lie group with the structure constants
subjected to conditions (7) and (8).  However, only for the linear odd
Poisson bracket realized in terms of the Grassmann variables and only in
the case when this bracket corresponds to the semi-simple Lie group,
there exists the Lie superalgebra (20)-(34) for the $\Delta$-like
operators of this bracket.

Note that in the case of the degenerate Cartan-Killing
metric tensor (9), relation (10) remains valid and we can construct only
two $\Delta$-like Grassmann-odd nilpotent operators: $\Delta_{-1}$ (24)
and $\Delta_{-3}$ (25), which satisfy the trivial anticommuting relation
$$
\{ \Delta_{-1}, \Delta_{-3} \} = 0\ .
$$
Note also that anticommuting relations for the operators
$$
\stackrel{i}\Delta_{-1} = {1\over\sqrt{2}} \Theta_\gamma
\stackrel{i}{{c_{\alpha\beta}}^\gamma}
\partial_{\Theta_\alpha} \partial_{\Theta_\beta}
\ ,\qquad (\stackrel{i}\Delta_{-1})^2 = 0\ ,
$$
corresponding to the Lie algebras with structure constants
$\stackrel{i}{{c_{\alpha\beta}}^\gamma}$ $(i = 1,2)$,
vanish provided that $\stackrel{i}{{c_{\alpha\beta}}^\gamma}$ satisfy
compatibility conditions \cite{blyt}
$$
\sum_{(\alpha\beta\gamma)} \stackrel{\{i}{{c_{\alpha\beta}}^\lambda}
\ \stackrel{k \}}{{c_{\lambda\gamma}}^\delta} = 0\ ,
$$
where $\{ ik \}$ denotes the symmetrization of the indices $i$ and $k$.

The Lie superalgebra (20)-(34), naturally connected with the linear odd
Poisson bracket (12), may be useful for the subsequent development of the
Batalin-Vilkovisky formalism for the quantization of gauge theories.
Indeed, very similar to (12) odd Poisson brackets on the Grassmann algebra
are used in a generalization \cite{g} of the triplectic formalism
\cite{bms} which is a covariant version of the $Sp(2)$-symmetric
quantization \cite{blt} of general gauge theories. We should therefore
expect that the Lie superalgebra (20)-(34), closely related with the
linear odd bracket (12), will also find the application for the further
development of the above-mentioned generalization of the triplectic
formalism. Let us note that the superalgebra (20)-(34) can also be used
in the theory of representations of the semi-simple Lie groups.

The author is sincerely thankful to V.D. Gershun, D.A. Leites and S.L.
Lyakhovich for useful discussions and is indebted to J.D. Stasheff for
illuminating remarks. The author wishes to thank J. Wess for kind
hospitality at the University of Munich where this work was completed.

This work was supported in part by the Ukrainian State Foundation
of Fundamental Researches, Grant No 2.5.1/54 and by Grant INTAS No 93-127
(Extension).

\end{document}